\documentclass[prl, twocolumn, superscriptaddress, reprint, aps]{revtex4-1}

\usepackage{empheq}
\usepackage{amsmath}  
\usepackage{amsfonts}
\usepackage{graphicx}   
\usepackage{hyperref}   
\usepackage{cleveref}

\usepackage{bm}
\usepackage{color}
\newcommand{\be}{\begin{enumerate}}
\newcommand{\ee}{\end{enumerate}}
\newcommand{\bi}{\begin{itemize}}
\newcommand{\ei}{\end{itemize}}

\newcommand{\xx}{\mathbf{x}}

\newcommand{\eq}{\begin{equation}}
\newcommand{\qe}{\end{equation}}
\newcommand{\bal}{\begin{align}}
\newcommand{\lab}{\end{align}}

\newcommand{\psidot}{\dot{\psi}}

\usepackage{array}
\usepackage{amssymb}

\begin{document}

\title{Amorphous topological insulators constructed from random point sets}
\author{Noah P. Mitchell}
\email{npmitchell@uchicago.edu}
\thanks{Corresponding author}
\affiliation{James Franck Institute and Department of Physics, University of Chicago, Chicago, IL 60637, USA}
\author{Lisa M. Nash}
\affiliation{James Franck Institute and Department of Physics, University of Chicago, Chicago, IL 60637, USA}
\author{Daniel Hexner}
\affiliation{James Franck Institute and Department of Physics, University of Chicago, Chicago, IL 60637, USA}
\author{Ari M. Turner}
\affiliation{Department of Physics, Israel Institute of Technology}
\author{William T. M. Irvine}
\email{wtmirvine@uchicago.edu}
\thanks{Corresponding author}
\affiliation{James Franck Institute and Department of Physics, University of Chicago, Chicago, IL 60637, USA}
\affiliation{Enrico Fermi Institute, The University of Chicago, Chicago, IL 60637, USA}

\maketitle
\textbf{
The discovery that the band structure of electronic insulators may be topologically non-trivial
has revealed distinct phases of electronic matter with novel properties~\cite{Kane_Quantum_2005, Haldane_Model_1988}. 
Recently, mechanical lattices have been found to have similarly rich structure in their phononic excitations~\cite{kane_topological_2013,prodan_topological_2009}, giving rise to protected uni-directional edge modes~\cite{nash_topological_2015,wang_topological_2015,susstrunk_observation_2015}. 
In all these cases, however, as well as in other topological metamaterials~\cite{sussman_topological_2016,kane_topological_2013}, the underlying structure  was finely tuned, be it through periodicity, quasi-periodicity or isostaticity. 
Here we show that amorphous Chern insulators can be readily constructed from arbitrary underlying structures, including  hyperuniform, jammed, quasi-crystalline, and uniformly random point sets. 
While our findings apply to mechanical and electronic systems alike, we focus on networks of interacting gyroscopes as a model system.
Local decorations control the topology of the vibrational spectrum, endowing amorphous structures with protected edge modes---with a chirality of choice.
Using a real-space generalization of the Chern number, we investigate the topology of our structures numerically, analytically and experimentally.
The robustness of our approach enables the topological design and self-assembly of non-crystalline topological metamaterials on the micro and macro scale.}

\begin{figure*}[ht]
\includegraphics[width=\textwidth]{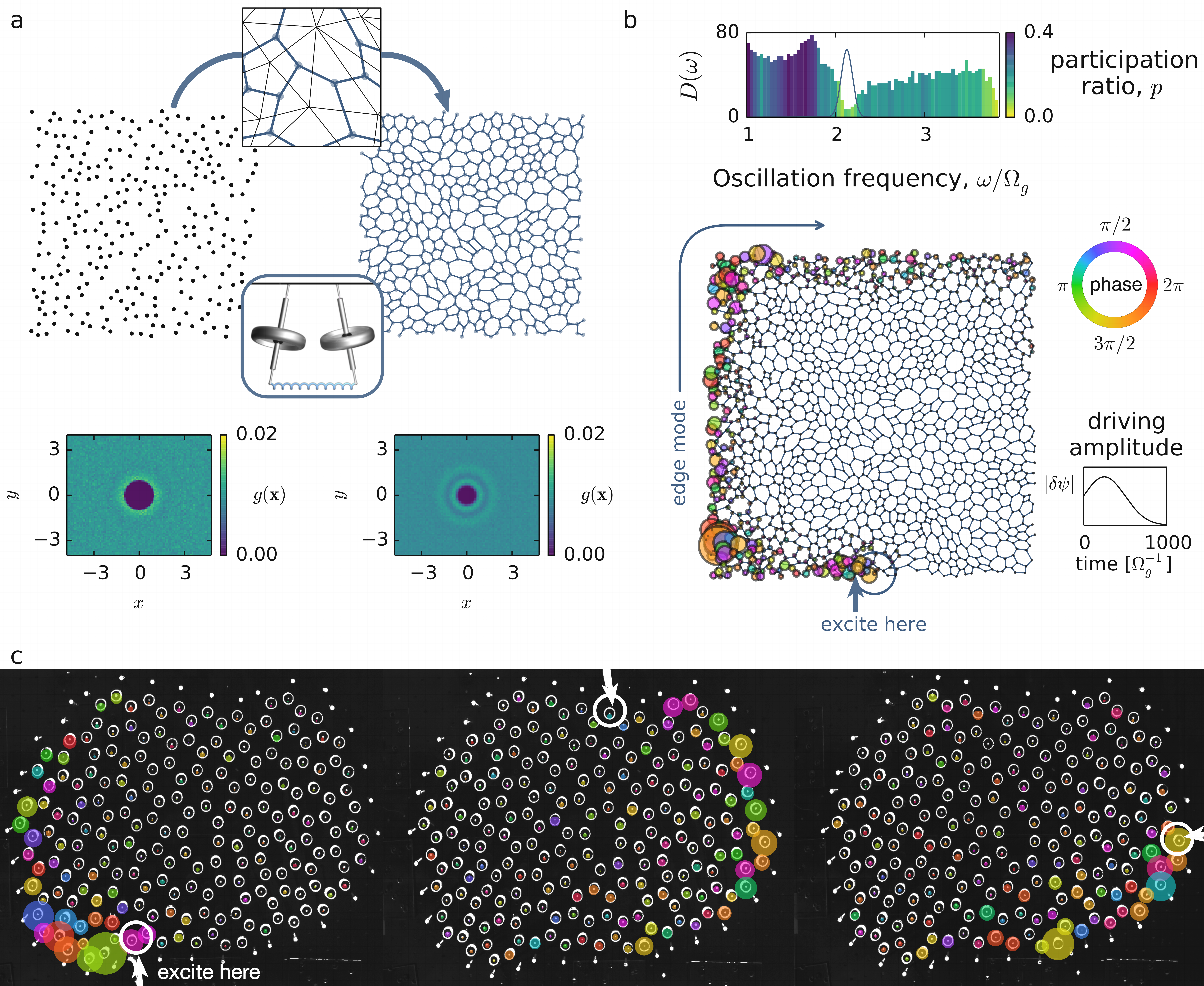}
\caption{
\textbf{Local structure gives rise to chiral edge modes $\mathbf{|}$}
\textbf{a} Voronoization of an amorphous structure, constructed by connecting adjacent centroids of a triangulation, preserves isotropy and lack of long-range order, here with a hyperuniform point set.
Two-point correlation functions $g(\xx)$ (below) reveal isotropic spatial structure for a system of $N \approx 3000$ particles. 
Spatial coordinates $x$ and $y$ are measured in units of the median bond length.
\textbf{b} Simulations reveal chiral edge modes in topological gyroscopic networks.
The localization of modes is probed by participation ratio, $p = {\left(\sum_i |\psi_i|^2 \right)^2} / {N \sum_i |\psi_i|^4} $, and the density of states is plotted as a function of normal mode oscillation frequency, in units of the gravitational precession frequency, $\Omega_g = \ell mg/I \omega $. 
The blue curve overlaying the density of states denotes the frequency of the driving excitation in the simulation. 
Here, the characteristic spring frequency, $\Omega_k = k \ell^2/ I \omega$ is chosen such that $\Omega_g = \Omega_k$.
The inset on the right shows the amplitude, $|\delta \psi|$, of the displacement for the single gyroscope which is shaken at a constant frequency.
\textbf{c}, An edge mode propagates clockwise in an amorphous experimental gyroscopic network. The motor- driven gyroscopes couple via a magnetic dipole-dipole interaction. 
Despite the nonlinear interaction and spinning speed disorder ($\sim 10 \%$), the edge mode appears, no matter where on the boundary the excitation is initialized.
}
\label{fig1}
\end{figure*}
Condensed matter science has traditionally focused on systems with underlying spatial order, as many natural systems spontaneously aggregate into crystals.
The behavior of amorphous materials, such as glasses, has remained more challenging~\cite{chaikin_principles_2000}.
In particular, our understanding of common concepts such as bandgaps and topological behavior in amorphous materials is still in its infancy when compared to crystalline counterparts.
This is not only a fundamental problem; advances in modern engineering, both of metamaterials and of quantum systems, has opened the door for the creation of materials with arbitrary structure, including amorphous materials.
This prompts a search for principles that can apply to a wide range of amorphous systems, from interacting atoms to mechanical metamaterials.

In the exploration of topological insulators, conceptual advances have proven to carry across between disparate physical realizations, from quantum systems~\cite{jotzu_experimental_2014}, to photonic waveguides~\cite{rechtsman_photonic_2013}, to acoustical resonators~\cite{yang_topological_2015,khanikaev_topologically_2015}, to hinged or geared mechanical structures~\cite{kane_topological_2013,Meeussen_geared_2016}.
One promising model system is a class of mechanical insulators consisting of gyroscopes suspended from a plate.
Appropriate crystalline arrangements of such gyroscopes break time-reversal symmetry, opening topological phononic band gaps and supporting robust chiral edge modes~\cite{nash_topological_2015,wang_topological_2015}.

Unlike trivial insulators, whose electronic states can be thought of as a sum of independent local insulating states, topological insulators require the existence of delocalized states in each nontrivial band and prevent a description in terms of a basis of localized Wannier states~\cite{thouless_wannier_1984,huo_current_1992,thonhauser_insulator/chern-insulator_2006}.  
It is natural, therefore, to assume that some regularity over long distances may be key to topological behavior, even if topological properties are robust to the addition of disorder. 
However, the extent to which spatial order needs to be built into the structure that gives rise to topological modes is unclear. 
We report a recipe for constructing amorphous arrangements of interacting gyroscopes---structurally more akin to a liquid than a solid---that naturally support topological phonon spectra.
By simply changing the local connectivity, we can tune the chirality of edge modes to be either clockwise or counter-clockwise, or even create both clockwise and counter-clockwise edge modes in a single material. 
This shows that topology, a nonlocal property, can naturally arise in materials for which the only design principle is the local connectivity.
Such a design principle lends itself to imperfect manufacturing and self-assembly.
Although our construction arises naturally in mechanical metamaterials, we show that it extends to electronic systems in the tight binding limit. 

\begin{figure}[ht]
\includegraphics[width=\columnwidth]{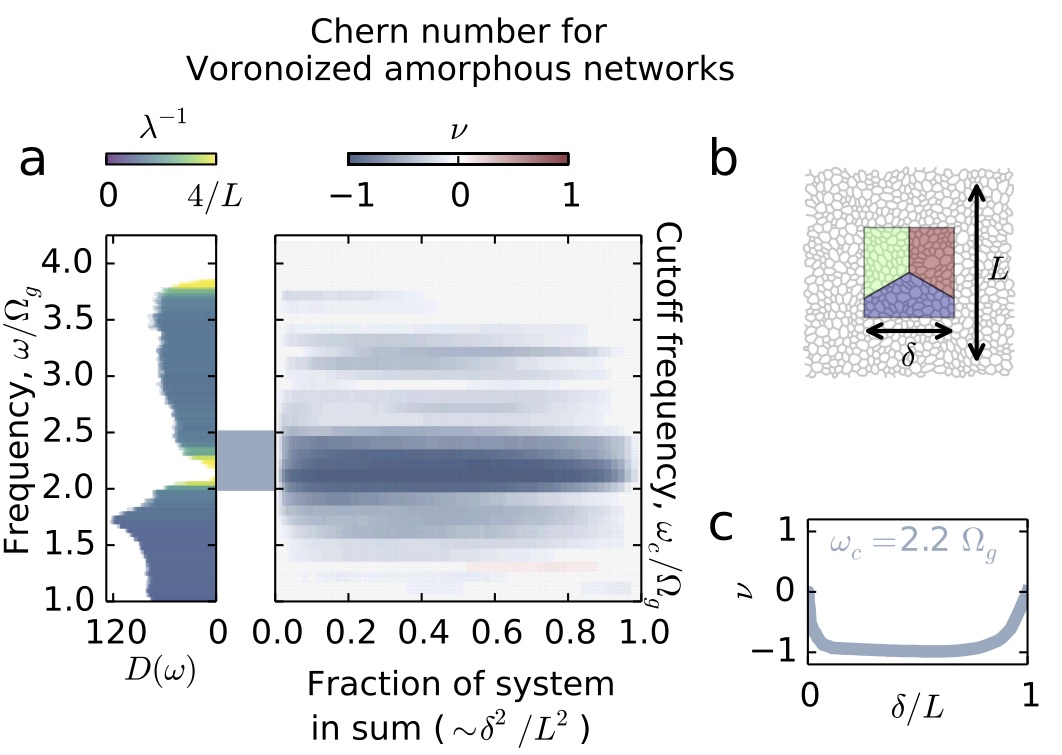}
\caption{\textbf{Chern number calculations confirm topological mobility gaps $\mathbf{|}$} 
\textbf{a}, The Chern number is computed for the band of frequencies above a cutoff frequency, $\omega_c$, using a real-space method. 
Once all modes in a band that carry Hall conductance are included, the Chern number converges to an integer value. 
On the left is an overlaid density of states $D(\omega)$ histogram for ten realizations of Voronoized hyperuniform point sets ($\sim 2000$ particles), with each mode colored by its inverse localization length, $\lambda^{-1}$. 
The topological mobility gaps remain in place and populated by highly localized states for all realizations.
(\textbf{b} and \textbf{c}), The computed Chern number converges once $\sim$20-40 gyroscopes are included in the summation region (red, green, blue regions panel \textbf{b}), and remains at an integer value until the summation region begins to enclose the sample boundary.
All networks have their precession and spring frequencies set to be equal ($\Omega_g = \Omega_k$).
}
\label{fig2}
\end{figure}

\begin{figure*}[ht]
\includegraphics[width=\textwidth]{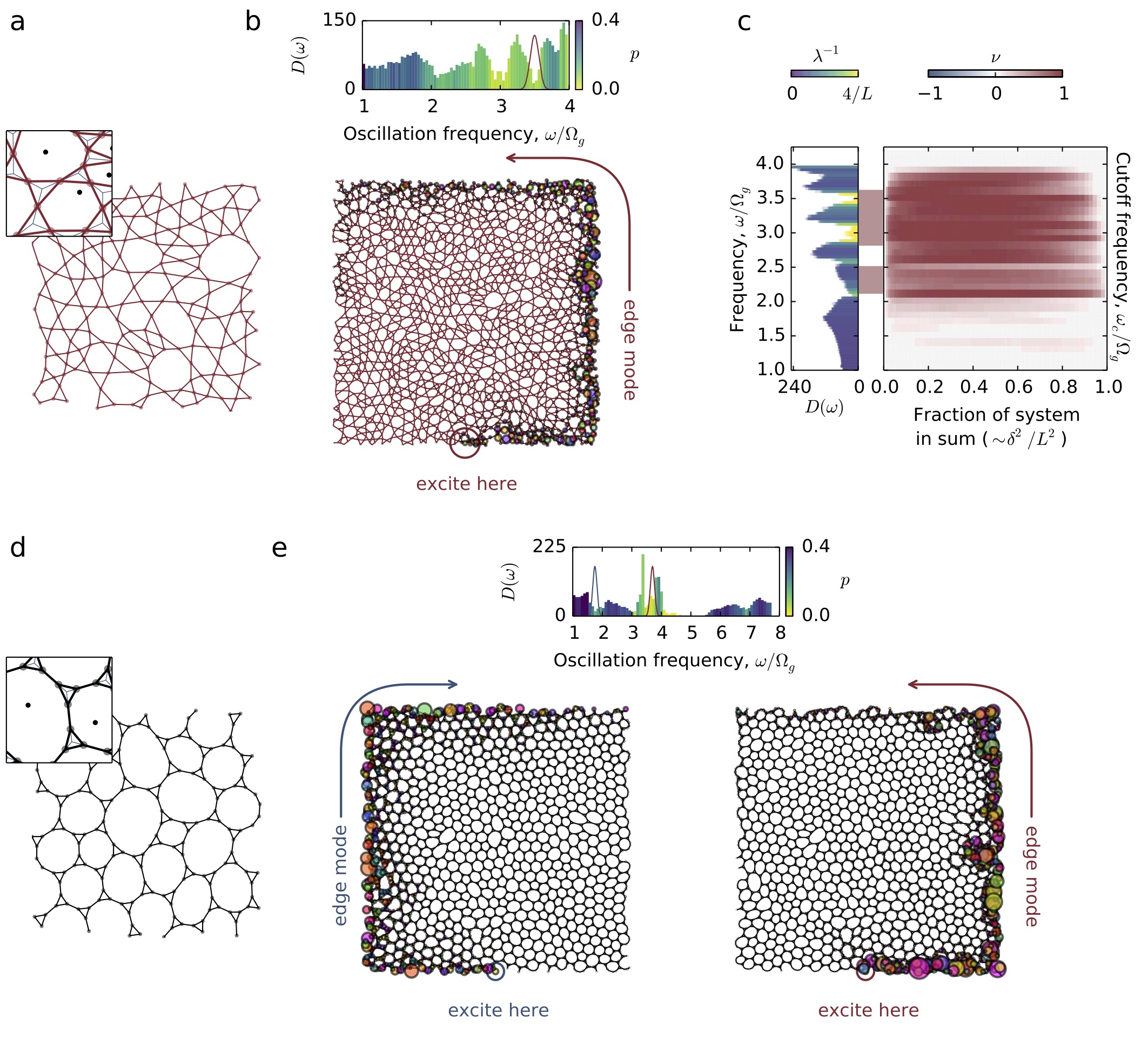}
\caption{
\textbf{Alternative local decorations allow control of the edge mode chirality $\mathbf{|}$}
(\textbf{a}-\textbf{c}), Kagomization of an arbitrary point set yields edge modes in gyroscopic networks with the opposite chirality to those in Voronoized networks.
(\textbf{d}-\textbf{e}), Another local decoration of the initial point set allows for multiple gaps with either chirality. The amorphous `spindle' network has two gaps with chiral edge modes: blue and red curves overlaying the density of states, $D(\omega)$, mark the excitation amplitude as a function of frequency for the two cases.
In (\textbf{b}-\textbf{c}), the spring frequency $\Omega_k = k \ell^2/ I \omega$ is set equal to the gravitational precession frequency, $\Omega_g$, while in (\textbf{e}), we chose $\Omega_k = 7 \Omega_g$ to broaden the lower (clockwise) mobility gap.
}
\label{fig3}
\end{figure*}

\begin{figure*}[ht]
\includegraphics[width=\textwidth]{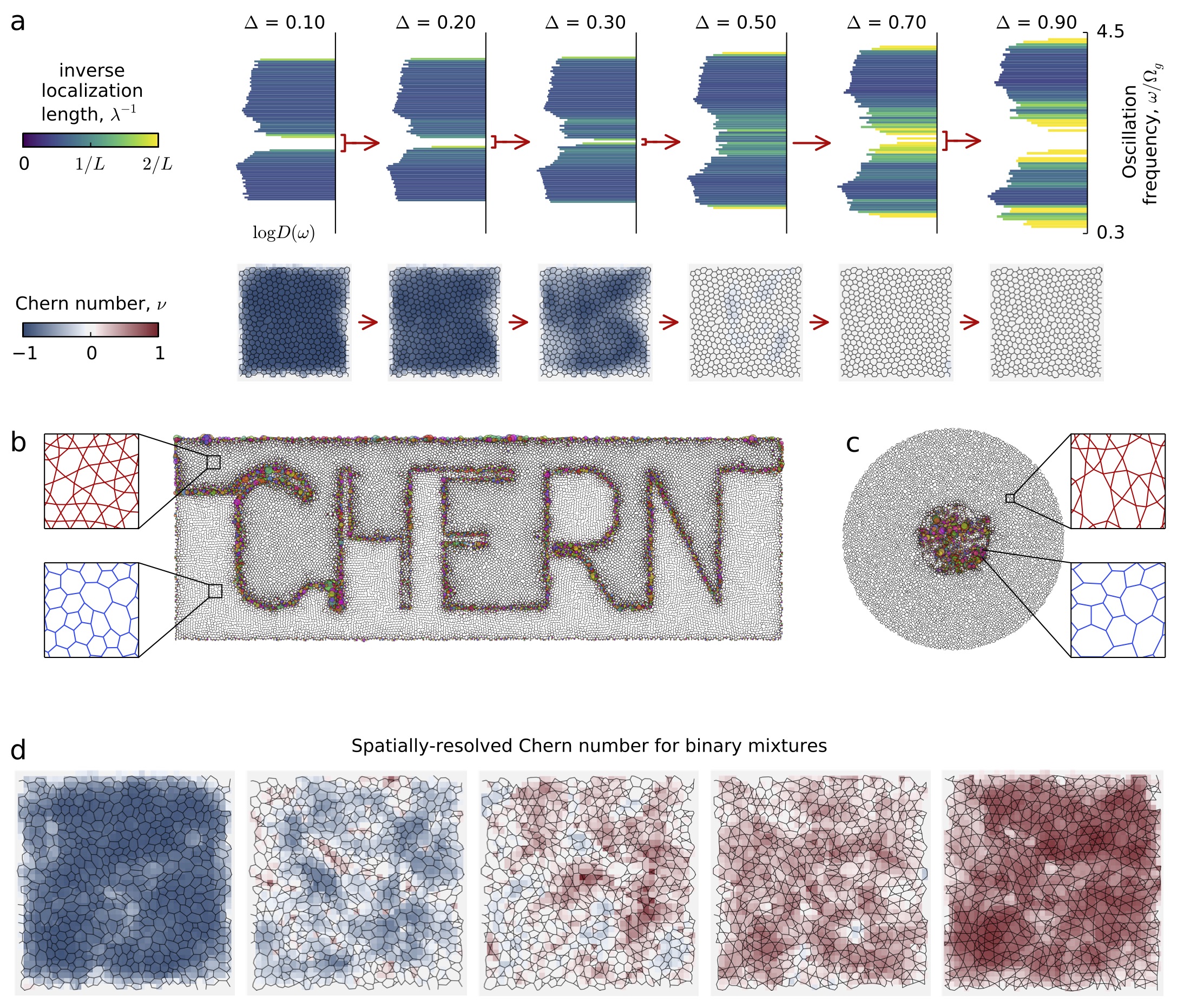}
\caption{
\textbf{Transition of a topological amorphous network to the trivial phase, and binary mixtures of Voronoized and kagomized networks $|$ } 
\textbf{a}, Locally breaking inversion symmetry by increasing and decreasing the precession frequencies of alternating gyroscopes competes with broken time reversal symmetry, triggering a transition to the trivial phase, with no edge modes. The precession frequency splitting, $\Delta$, is tuned so that $\Omega_g^{A} =\Omega_k (1+\Delta)$ and $\Omega_g^{B} = \Omega_k (1-\Delta)$. 
\textbf{b}, Edge modes are localized at the interfaces between kagomized and Voronoized networks, permitting sinuous channels for the propagation of unidirectional phonons. 
\textbf{c}, Excitations of a Voronoized region nested inside a kagomized network remain confined when the excitation frequency is in a mobility gap unique to the kagomized network.
\textbf{d}, When kagomized elements are randomly mixed into a Voronoized network, the sign of the local, spatially-resolved Chern calculation is determined by the local geometry, with excitations in a mobility gap biased toward the interface of the two clusters.
}
\label{fig_binary}
\end{figure*}

The first local construction we consider is shown in Figure~\ref{fig1}a.
Starting from an arbitrary point set, a natural way to form a network is to generate a Voronoi tessellation, either via the Wigner-Seitz construction or by connecting centroids of a triangulation~\cite{florescu_designer_2009}. 
Treating the edges of the cells as bonds and placing gyroscopes at the vertices leads to a network reminiscent of `topological disorder' in electronic systems~\cite{weaire_electronic_1971}.
A range of frequencies arises in which all modes are tightly localized, and this frequency region overlaps with the corresponding band gap of the honeycomb lattice.
Crucially, we find that gyroscope-and-spring networks constructed in this way from arbitrary initial point sets invariably have such a mobility gap in a frequency range determined by the strength of the gravitational pinning and spring interactions.

Our networks are reminiscent of `topologically disordered' electronic systems~\cite{weaire_electronic_1971}. 
In these systems, a central characteristic is that the local density of states as a function of frequency is predictive of the global density of states. 
Specifically, band gaps or mobility gaps are preserved~\cite{weaire_electronic_1971,weaire_existence_1971,haydock_electronic_1972}. 
Interestingly, we find that, even in the presence of band topology, averaging the local density of states over mesoscopic patches ($\sim 10$ gyroscopes) reproduces the essential features of the global density of states as a function of frequency.
Furthermore, we find that inserting mesoscopic patches of our structures into a variety of other dissimilar networks (see Supplementary Information Figures 8-10) does not significantly disrupt the averaged local density of states of the patch.

Crucially, we find that our structures show hallmarks of non-trivial topology. 
When the system is cut to a finite size, modes confined to the edge populate the mobility gap, mixed in with localized states.
As shown in the direct simulations of Figure~\ref{fig1} and Supplementary Videos 1 and 2, shaking a gyroscope on the boundary results in chiral waves that bear all the hallmarks of protected edge states (robustness to disorder and absence of back-scattering).

An experimental realization can be readily constructed from gyroscopes interacting magnetically, as seen from below in Figure~\ref{fig1}c.
Like in~\cite{nash_topological_2015}, these gyroscopes consist of 3D-printed units encasing DC motors which interact via magnetic repulsion. 
Probing the edge of this system immediately generates a chiral wave packet localized to the boundary, confirming 
that this class of topological material is physically realizable and robust (Figure~\ref{fig1}c and Supplementary Videos 3 and 4).

This behavior begs for a topological characterization, even though it might be surprising that topology can emerge from such a local construction.
The existence of chiral edge states in an energy gap is guaranteed if an invariant known as the Chern number is nonzero, and the direction of the chiral waves is given by its sign.
Although the Chern number was originally defined in momentum space, several generalizations have been constructed in coordinate space in order to accommodate disorder in crystalline electronic materials~\cite{Kitaev_anyons_2006,prodan_non-commutative_2010,bianco_mapping_2011}.
In these methods, information about the system's vibrations above a cutoff frequency, $\omega_c$, is carried by the projection operator, $P$.
Each element $P_{ij}$ measures the response of gyroscope $j$ to excitations of gyroscope $i$ within a prescribed range (band) of frequencies. 

According to one such method, proposed in~\cite{Kitaev_anyons_2006}, a subset of the system is divided into three parts and labeled in a counterclockwise fashion (red, green, and blue regions in Figure~\ref{fig2}).
These regions are then used to index components of an antisymmetric product of projection operators:
\begin{equation}\label{eq_kitaev}
\nu(P) = 12 \pi i \sum_{j \in A} \sum_{k \in B} \sum_{l \in C}  \left(
P_{jk} P_{kl} P_{lj} - P_{jl} P_{lk} P_{kj} \right).
\end{equation}
The sum of such elements converges to the Chern number of the band above a chosen cutoff frequency, $\omega_c$, when the summation region has enclosed many gyroscopes (see Supplementary Videos 5 and 6 and Supplementary Information).

Equation~\ref{eq_kitaev} can be understood as a form of charge polarization in the response of an electronic material to a locally applied magnetic field.  
Applying a magnetic field to a small region of a material induces an electromotive force winding around the site of application. 
If the material is a trivial insulator, any changes in charge density there arise from local charge re-arrangements, which result in no accumulation of charge. 
By contrast, a topological  electronic system has a Hall conductivity determined by the Chern number. 
As a result, a net current will flow perpendicular to the electromotive force, inducing a net-nonzero charge concentrated at the magnetic field site, compensated by charge on the boundary. 
As we show in the Supplementary Information, the amount of local charging is proportional to the applied field, and the proportionality constant is the Chern number of Equation~\ref{eq_kitaev}.

Figure~\ref{fig2}a shows the results of Equation~\ref{eq_kitaev} computed for the Voronoized networks. 
As the cutoff frequency for the projector is varied (here it is lowered from $4\Omega_g$), the computed Chern number converges to $\nu = -1$ when all extended states in the top band lie above the cutoff frequency, confirming that the modes observed in Figure~\ref{fig1}b and c are topological in origin and predicting their direction. 
The Chern number remains at its value of $\nu=-1$ for a broad range of frequencies in which any existing states are localized, and thus do not contribute to the Chern number. The Chern number returns to zero once more conductance-carrying extended states are included in the calculation.

Having established this connection, 
we now discuss how the Chern number can be controlled.
In particular, we show that by considering alternative decorations of the same initial point set, it is possible to flip the chirality of the edge modes or even provide multiple gaps with differing  chirality.
One possible construction arises naturally from joining neighboring points in the original point set, leading to a Delaunay triangulation.
As shown in Supplementary Video 7, such networks show no gaps and no topology, suggesting that the local geometry dictated by Voronoization is responsible for its emergent topology.
A clue can be found by noting that the Voronoized networks are locally akin to a honeycomb lattice.
The honeycomb is the simplest lattice with more than one site per unit cell, a necessary condition for supporting a band gap in a lattice.
Moreover, this lattice was previously found to be topological with the same Chern number~\cite{nash_topological_2015}. 

Building on this insight, we introduce a second decoration, which we dub `kagomization,' shown in Figure~\ref{fig3}a. 
If applied to a triangular point set, Voronoization produces a honeycomb lattice and kagomization produces a kagome lattice, the simplest lattice with three sites per unit cell, which we have found to produce $\nu=+1$ gyroscopic metamaterials. 
Proceeding as with the Voronoized network case (Figures~\ref{fig3}b and c), we find the presence of topologically protected modes with opposite direction and the corresponding opposite Chern number in the the band structure (see Supplementary Video 8).
Other local constructions, such as the `spindle' networks in Figure~\ref{fig3}d-e provide multiple mobility gaps, each with a different edge mode chirality, offering a transmission direction tuned by frequency (see Supplementary Video 9).

One might think there could be a mapping from the geometry of each vertex to the chirality of the edge modes. 
However, taken together, our Voronoized, kagomized, and spindle networks demonstrate that simply counting nearest neighbors is not sufficient to determine the topology: 
a description beyond nearest neighbors is required (see also Supplementary Information Figure 29). 
On the other hand, we are able to change the Chern number of a structure via local decorations.
To uncover the extent to which a network's topology is stored locally, consider the projection operator, $P$.  
The projector value $P_{ij}$ measures the vibrational correlation between gyroscope $j$ and gyroscope $i$ when considering all modes above a cutoff frequency. 
By explicitly computing its magnitude in our networks, we find that the magnitude of $P_{ij}$ falls off exponentially with distance (Supplementary Information Figure 11). 
Remarkably, explicitly cutting out a section of the network and embedding it in a network with a different spectrum results in only a slight change to the local projector values ($< 2\%$) (see Supplementary Information Figure 12). 
Since the Chern number is built from these projector elements, it then follows that the local structure of the gyroscope network, combined with some homogeneity of this local structure across the lattice, is all that is needed to determine the Chern number (c.f. Supplementary Video 10 and the Supplementary Information section entitled `Sign of Chern number from network geometry'). 

This situation is reminiscent of electronic glasses in which the local binding structure gives rise to a local `gap.' 
Under weak assumptions of homogeneity, this gap can be shown to extend to the whole system~\cite{weaire_existence_1971,weaire_electronic_1971}.
The case with topology is similar: the next-nearest neighbor angles in a network's cell open a local `gap' by breaking time reversal symmetry.

For amorphous networks, we make the correspondence between the bulk topological invariant and the edge states on the boundaries by considering a gyroscopic sample shaped into an annulus (c.f.~\cite{laughlin_quantized_1981,halperin_quantized_1982,fu_topological_2007}).
Adiabatically tuning the interactions between pairs of gyroscopes along a radial cut (by adding a fixture to one gyro from each pair) pumps each edge mode into a neighboring mode, as shown in Supplementary Video 11 and in the Supplementary Information section `Spectral flow through adiabatic pumping'.
If we consider all states below a gap cutoff frequency, the process---which mimics the effect of threading a magnetic field through the centre of an annulus in an electronic system---trades one state localized on the outer boundary for an extra state on the inner boundary of the annulus, which we connect to the real-space Chern number (Equation~\ref{eq_kitaev}) in the Supplementary Information. 

As in the lattice case, a mobility gap becomes topological due to time reversal symmetry breaking: bond angles in these networks are not multiples of $\pi/2$ (c.f.~\cite{nash_topological_2015}).
We can probe this mechanism by eliminating a gap's topology. 
Alternating the gravitational precession frequency, $\Omega_g$, of neighboring gyroscopes in a network mimics the breaking of inversion symmetry on a local scale, an effect which competes against the time reversal gap opening mechanism.
When the precession frequency difference between sites is large enough, this competing mechanism eliminates edge modes, triggering a transition to a $\nu =0$ mobility gap, shown in Figure~\ref{fig_binary}a and Supplementary Videos 12 and 13.

Equipped with these insights, we can easily engineer networks which are heterogeneous mixtures of multiple local configurations.
Figure~\ref{fig_binary}b-d highlight some results of combining Voronoized and kagomized networks or encapsulating one within another.
Because the Voronoized and kagomized networks share a mobility gap, excitations are localized to their interface, offering a method of creating robust unidirectional waveguides, such as the sinuous waveguide shown in Figure~\ref{fig_binary}b and Supplementary Video 14. 
Figure~\ref{fig_binary}c demonstrates that additional topological mobility gaps at higher frequency in the kagomized network allow bulk excitations to be confined to an encapsulated Voronoized region (see also Supplementary Video 15).
Random mixtures of the two decorations, shown in Figure~\ref{fig_binary}d, demonstrate heterogeneous local Chern number measurements (red for $\nu = +1$ and blue for $\nu = -1$), with mobility-gap excitations biased toward the interfaces between red and blue regions (see Supplementary Information and Supplementary Video 16).

As our networks are structurally akin to liquids, they support topological modes in the absence of long range spatial order.
The details of the underlying point set are not essential, and neither are the details of the local Voronoization or kagomization procedures. 
We verified this by replacing the centroidal construction~\cite{florescu_designer_2009} with a Wigner-Seitz construction (see Supplementary Information Figures 24-26 for a comparison). 
Beyond mechanical materials, we find similar results in electronic tight binding models of amorphous networks, underscoring the generality of the finding (Figure 32 of the Supplementary Information).

This study demonstrates that local interactions and local geometric arrangements are sufficient to generate chiral edge modes, promising new avenues for engineering topological mechanical metamaterials generated via imperfect self-assembly processes.
Such self-assembled materials could be constructed, for instance, with micron-scale spinning magnetic particles.
Since our methods bear substantial resemblance to tight-binding models, our results also find direct application not only to electronic materials, as we have demonstrated, but also to photonic topological insulators~\cite{rechtsman_photonic_2013}, acoustic resonators~\cite{yang_topological_2015,khanikaev_topologically_2015}, and coupled circuits~\cite{ningyuan_time-_2015}.

\textbf{Acknowledgements:} We thank Michael Levin, Charlie Kane, and Emil Prodan for useful discussions. 
This work was primarily supported by the University of Chicago Materials Research Science and Engineering Center, which is funded by National Science Foundation under award number DMR-1420709. Additional support was provided by the  Packard Foundation. The Chicago MRSEC (U.S. NSF grant DMR 1420709) is also gratefully acknowledged for access to its shared experimental facilities. 
This work was also supported by NSF EFRI NewLAW grant 1741685.

 \textbf{Note added in proof:}
 After the submission of this work, we became aware of a concurrent theoretical study of the existence of amorphous electronic topological insulators in two and three dimensions~\cite{agarwala_topological_2017}.

\textbf{Contributions:} 
WTMI and NPM designed research. WTMI and AMT supervised research. NPM and LMN performed the simulations and experiments.  AMT, NPM, DH, and WTMI performed analytical calculations. DH contributed numerical tools.  NPM and WTMI wrote the manuscript. All authors interpreted the results and reviewed the manuscript.

\textbf{Data Availability:} The data that support the plots within this paper and other findings of this study are available from the corresponding authors upon request.

\textbf{Methods:}
For small displacements $(\delta X, \delta Y)$ of the pivot points from vertical alignment with the centers of mass, the equation of motion for a gyroscope takes the form
\eq \label{simulation}
\left( 
\begin{array}{c}
 \dot{\delta X_i}\\
\dot{\delta Y_i}
\end{array}
\right)
 \approx
\Omega_g \left( \begin{array}{c}
\delta Y _i  \\
- \delta X_i
 \end{array} \right)
 +
 \frac{\Omega_k}{k}
\sum_j^{\textrm{NN}}
\left( \begin{array}{c}
-F_{ij, Y} \\ F_{ij, X}
\end{array} \right),
\qe
where $\Omega_k = k \ell^2/ (I \omega)$, and $\Omega_g = \ell mg/(I \omega) $, while $k$ is the spring constant, $\ell$ is the length of the pendulum from the pivot to the center of mass, $I$ is the moment of inertia along the spinning axis, $\omega$ is the angular spinning frequency, $m$ is the mass of the gyroscope, and $g$ is the gravitational acceleration.
The simulations evolve Equation~\ref{simulation} using a Runge-Kutta fourth-order explicit method run on a GPU using OpenCL. 

To obtain spectra and normal modes, note that Equation~\ref{simulation} defines the entries for a system's dynamical matrix, $ \mathbf{D}$, such that
\eq
\dot{\vec{\psi}} = \mathbf{D} \vec{\psi},
\qe
where the components of $\vec{\psi}$ contain information on the displacements of the gyroscopes. 
In order to map to a tight binding model, it is useful to write the displacement of the $i^{th}$ gyroscope as $\psi_i \equiv \delta X_i + i \delta Y_i$ and note that the eigenvalue problem gives pairs of eigenvalues $\pm\omega$, so that
\eq
\psi_i = L_i e^{i\omega t} + \bar{R}_i e^{-i \omega t} 
\equiv \psi^L_i + \bar{\psi}^R_i.
\qe
Then
\begin{multline}
i \left(
\begin{array}{c}
 \psidot^L_i \\
\dot{\psi}^R_i
\end{array}
\right) 
 = \Omega_g \left( \begin{array}{c}
-\psi^L_i \\
\psi^R_i 
\end{array}
\right)
\\
 - \frac{ \Omega_k}{2}
\sum_{j \in \textrm{NN}(i)} \left(
\begin{array}{c}
\psi_{i}^L -\psi_{j}^L 
+ e^{i2\theta_{ij}}(\psi_i^R - \psi_j^R) \\
-\psi_{i}^R +\psi_{j}^R 
- e^{-i2\theta_{ij}}(\psi_i^L - \psi_j^L ) 
\end{array}
\right).
\end{multline}

The experiment in Figure 1c shows a chiral wave packet localized to the boundary in an experimental realization of 195 gyroscopes in a Voronoized network based on a point set generated from a jammed packing. 
Each gyroscope consists of a spinning motor ($\sim 150$ Hz) housed in a 3D-printed enclosure (as in~\cite{nash_topological_2015}), and each gyroscope is suspended from an acrylic plate by a spring, an attachment method which was found to reduce damping. 

To establish the equivalence of Equation~\ref{eq_kitaev} and the response to a point-like magnetic field, we study the effect of a perturbation on the projection operator $P_{ij} \equiv \sum_{n\in \mathrm{band}} \chi_n(i)\chi_n^*(j)$, where $\chi$ are the perturbed wavefunctions.
In the Supplementary Information, we link the change in the diagonal elements to the charge accumulation near an applied magnetic field:
\begin{equation}\label{IQHE_Bfield}
\Delta \rho= \nu \frac{e^2}{h} B_z,
\end{equation} 
where $\Delta \rho$ is the change in charge density where the magnetic field is applied, $B_z$ is the magnetic field normal to the sample, and $\nu$ is the Chern number of the occupied bands when the sample is periodic.

Having shown that the Chern sum (Equation~\ref{eq_kitaev}) is equal to the charge accumulated when a quantum of magnetic flux is inserted, we can establish a correspondence between the bulk invariant and edge modes on the boundary by introducing a hole at the site of insertion.
The real-space Chern number is then equal to the number of edge states that accumulate on the inner boundary as an effective magnetic flux is introduced through the hole~\cite{laughlin_quantized_1981,halperin_quantized_1982,fu_topological_2007}.
The effective magnetic flux is manifest as a phase shift in the interactions for any loop of spring connections that encloses the hole.
We construct this phase shift by altering the subset of the nearest-neighbor interactions that traverse a cut of the annulus, such that the force of one gyro on its neighbor across the cut is altered by a rotation
\begin{equation}
F \sim \psi_i -\psi_j \rightarrow 
\psi_i - \psi_j e^{i\theta_{\textrm{twist}}}.
\end{equation}
In the Supplementary Information, we propose a concrete picture of how this could be built in an experiment by attaching an extensible ring to a small number of gyroscopes.

To see topological robustness in a simpler situation, we find similar behavior in an amorphous electronic tight binding model using the model Hamiltonian
\eq\label{electronic_model}
H = -t_1 \sum_{\langle ij \rangle} c_i^\dagger c_j - t_2  \sum_{\langle \langle ij \rangle \rangle}  
e^{-i\phi_{ij}}
c_{i}^\dagger c_j,
\qe
where $\langle ij \rangle$ denotes nearest neighbors $ij$ and $\langle \langle ij \rangle \rangle$ denotes pairs of next-nearest neighbors (NNN).
The parameter $t_2$ tunes the strength of all NNN hoppings, and $\phi_{ij}$ controls the degree to which the hopping $i\rightarrow j$ breaks time reversal symmetry (by tuning the imaginary term).
As shown in the Supplementary Information, topological edge modes arise in amorphous tight binding lattices, whether the NNN hopping is uniform or bond angle-dependent.


\end{document}